\documentstyle[12pt,epsfig]{article}

\begin{document}

\def\Journal#1#2#3#4{{#1} {\bf #2}, #3 (#4)}
\def\NIM{\em Nucl. Instrum. Methods}
\def\NIMA{{\em Nucl. Instrum. Methods} A}
\def\NPB{{\em Nucl. Phys.} B}
\def\PLB{{\em Phys. Lett.}  B}
\def\PRL{\em Phys. Rev. Lett.}
\def\PRD{{\em Phys. Rev.} D}
\def\ZPC{{\em Z. Phys.} C}

  \large

\title{
  \Large
\bf
\boldmath 
Mixing and $CP$-Violation Measurements of $B^0$ Mesons
from the Tevatron Collider}

\author{
  \large
G. Bauer\\ 
\lowercase{(representing the \uppercase{CDF} and 
                             \uppercase{D{\O} C}ollaborations)\\}
Massachusetts Institute of Technology, Cambridge, MA 02139, USA\\
E-mail: bauerg@fnal.gov
$\,$\\
$\,$\\
\date{}
}

\maketitle

\vspace*{-11.5cm}
{\hspace*{8.0cm} \large
\raggedleft
 FERMILAB-CONF-99/042-E\\
}
\vspace*{8.5cm}

\begin{abstract}
  \normalsize
$B^0$ mixing measurements from the Tevatron Run I data are reported.
These include time-integrated measurements of the average mixing
parameter $\overline{\chi}$, six time-dependent oscillation measurements
of $\Delta m_d$, 
and a time-dependent limit on $\Delta m_s$. 
Such measurements provide constraints on
CKM matrix elements.
A sample of $B^0/\overline{B}{^0} \rightarrow J/\psi K^0_S$ decays
is used to directly measure the $CP$-violation parameter $\sin(2\beta)$.
This value agrees well with indirect constraints on the CKM matrix.
\end{abstract}

\vspace*{0.2cm}
\section{
  \Large
 $B$-Physics  and 
       the CKM Matrix}\label{subsec:Tev}
A major objective in the study of bottom hadrons is determining 
the elements of the Cabbibo-Kobayashi-Maskawa (CKM) matrix~\cite{CKM},
and to stringently test its adequacy.
This matrix transforms the flavor eigenstates of quarks into their mass
eigenstates, which are not the same in the Standard Model (SM).
A convenient parameterization in powers of the Cabibbo
angle ($\lambda \equiv \sin \,\theta_C = |V_{us}|$)
is due to Wolfenstein~\cite{Wolf}:
 $$  V_{CKM} \equiv
 \left( \begin{array}{ccc}
   V_{ud} &  V_{us} &  V_{ub} \\
   V_{cd} &  V_{cs} &  V_{cb} \\
   V_{td} &  V_{ts} &  V_{tb} \\
        \end{array} \right)
   =
       \left( \begin{array}{ccc}
  1-\frac{\lambda^2}{2}  &      \lambda           &
A\lambda^3(\rho-i\eta)  \\
    { -\lambda}  &  1-\frac{\lambda^2}{2} &  { A\lambda^2} \\
{ A\lambda^3(1-\rho-i\eta)} &  -A\lambda^2     &  { 1}
\end{array} \right)
  + O(\lambda^4).
$$
The imaginary term $\eta$ was conjectured by Kobayashi and Maskawa
to be the source of $CP$ violation, which has been an outstanding issue for
the last 35 years.

Constraints on the CKM matrix from the $b$-sector initially came 
from lifetime and branching ratio measurements
in the early '80's. In 1986, a new window was opened 
by the observation of $B^0$-$\overline{B}{^0}$ mixing
in an unresolved mixture of $B^0_d$ and $B^0_s$ by UA1~\cite{UA1}
in $\bar{p}p$ collisions,
and subsequently for pure $B^0_d$'s by ARGUS~\cite{ARGUS}
at the $\Upsilon(4S)$.
Through mixing, one gains access to the $t$ CKM elements,
an important
consideration given the limitations of direct top studies.

\begin{figure}[t]
{\raggedright
  \centering
  \epsfxsize=20pc 
  \epsfbox{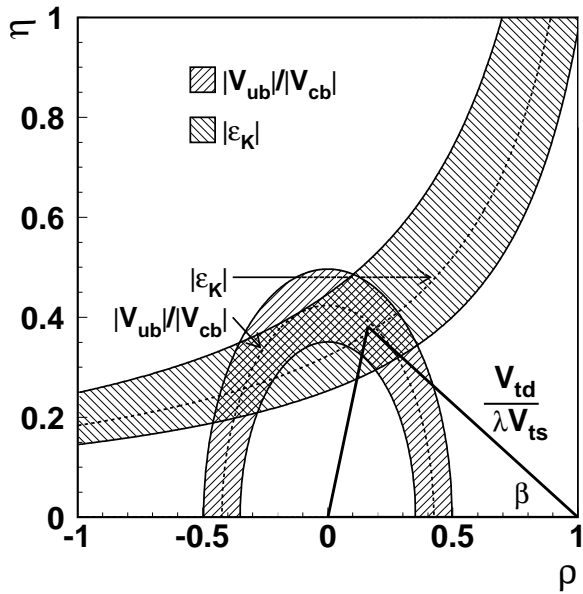}
  \caption{  
The CKM unitarity triangle 
  along with constraints derived from
  $CP$ violation in $K^0$'s ($\varepsilon_K$), 
  and the rate of charmless $B$ decays ($V_{ub}/V_{cb}$)
(derived from  Mele\protect~\protect\cite{Mele}).
Bands indicate ``$1\sigma$'' allowed regions.
The angle 
$\beta =  arg(-V^{\,}_{cd}V^*_{cb}/V^{\,}_{td}V^*_{tb})
= \arctan(\eta/[1-\rho])$.
}
\label{fig:Unitarity1} 
}
\end{figure}

Global fits to experimental data constrain
the four parameters of the CKM~\cite{Mele,Ali}, 
with $\lambda$ and $A$ already known quite well. 
Constraining $\eta$ and $\rho$ has been
the recent focus of $B$-physics.
One of the unitarity constraints 
($V_{tb}^*V^{\,}_{td} + V_{cb}^*V^{\,}_{cd}+V_{ub}^*V^{\,}_{ud} = 0$), 
is graphically represented in Fig.~\ref{fig:Unitarity1}
as a triangle in the complex $\rho$-$\eta$ plane,
with the apex at ($\rho$,$\eta$).
Its base is of unit length,
leaving three angles and two sides that may be measured.
$B^0$-$\overline{B}{^0}$ mixing constrains the right leg
($\propto\!V_{td}/V_{ts}$),
and $CP$ violation in 
$B^0/\overline{B}{^0} \rightarrow J/\psi K^0_S$ decays
determines the angle $\beta$.

Mixing studies of $B^0$ mesons have greatly advanced in the '90's,
and we are entering a new stage at the close of the millennium
with the advent of $CP$-violation measurements in $B^0$ mesons. 
The contributions from the Fermilab Tevatron Collider program 
to these efforts from Run I (1992-6) data are discussed. 
The two collider experiments, CDF~\cite{CDF} and D{\O}~\cite{D0}, 
are well known, 
and their descriptions are not repeated here.

\section{
  \Large
 $B^0$-$\overline{B}{^0}$ Mixing Measurements}
\subsection{
  \Large
$B$ Mixing and Flavor Tagging}
Like the $K^0$-$\overline{K}{^0}$ system,
2nd order weak ``box'' diagrams result in oscillations
between $B^0$ and $\overline{B}{^0}$ mesons. 
The frequency is the mass difference
($\Delta m = m_H - m_L$) between the heavy/light mass
eigenstates $B^0_H$ and $B^0_L$.
The dominant effect arises from diagrams with virtual top quarks,
with $\Delta m_q \sim m^2_t F(m^2_t/m^2_W) |V^*_{tb}V_{tq}|$
for $B^0_q$ mesons.
Thus, $\Delta m_q$ constrains the CKM element
relating transitions between top and the light quark $q$
composing the $B^0_q$, and relates to the right leg of
the unitarity triangle (Fig.~\ref{fig:Unitarity1}).
The probability that an initially pure $B^0$ state
decays as a $\overline{B}{^0}$ 
(and {\it vice versa}) at proper time $t$ is
${\cal P}_{Mix}(t) = [1 -\cos(\Delta m t)]e^{-t/\tau}/2\tau$.
The asymmetry between the mixed (${\cal P}_{Mix}$) 
and unmixed (${\cal P}_{Un}$) states is therefore
\begin{equation}
{\cal A}_0(t) \equiv \frac{P_{Un}(t)-P_{Mix}(t)}{P_{Un}(t)+P_{Mix}(t)}
= \cos(\Delta m_q t).
\label{eq:MixAsym}
\label{eq:true_asym}
\end{equation}

Observing $B^0$ mixing is predicated upon determining 
the $b$ ``flavor''---whether the $B$ is composed 
of a $b$ or a $\bar{b}$ quark---at the times of production
and decay.
The decay flavor is usually known from the $B$
reconstruction.
More problematic is tagging the initial flavor.
If this is correct with probability $P_0$,
then the observed asymmetry is attenuated by the ``dilution''
${\cal D}_0 = 2P_0 -1$, {\it i.e.} ${\cal A}_{Obs} = {\cal D}_0 \cos(\Delta m_q t)$.
A tagger 
with efficiency $\epsilon$ yields an error
on the asymmetry which scales as $1/\sqrt{\epsilon {\cal D}^2_0 N}$ for 
$N$ background-free mesons, and thus
$\epsilon {\cal D}^2_0$ measures the tagger's effective power.

\subsection{
  \Large
Time-Integrated Mixing Measurements}
\label{sec:timeint}
It is not necessary to measure the proper decay time
to observe mixing,
since
\begin{equation}
\chi \equiv \int^\infty_0{\cal P}_{Mix}(t) dt = \frac{x^2}{2(1+x^2)}, 
            \;\; {\rm with}\;\;
            x \equiv \frac{\Delta m}{\Gamma},
\label{eq:chi}
\end{equation}
is nonzero.
At the Tevatron both $B^0_d$ and $B^0_s$ are produced, and unless one explicitly 
identifies the $B$-species one measures
an average $\overline{\chi} \equiv f_d\chi_d + f_s\chi_s$,
for fractions $f_d$
and $f_s$ of the $B^0_d$ and $B^0_s$ contributions.

CDF and D{\O} have measured  $\overline{\chi}$ using
dileptons, where the leptons identify both
the $B$ and tag its flavor. Like-sign pairs indicate that one
$b$-hadron has mixed.
In $10$ pb$^{-1}$ of dimuon triggers ($p_T^\mu>3$ GeV/$c$)
D{\O} found 59 like-sign (LS) and 113 unlike (US)  pairs. 
The ratio $LS/US = 0.43 \pm 0.07 (stat.)\pm 0.05(syst.)$ 
is used in conjunction with models of other processes 
($b\rightarrow c \rightarrow \ell^+$ sequential decays, $c\bar{c}$,
fake leptons,\ldots etc.) to extract
 $\overline{\chi} = 0.09 \pm 0.04 \pm 0.03$~\cite{D0Mix}.
Similarly, CDF has used $20$ pb$^{-1}$ of dimuons 
to obtain
 $\overline{\chi} = 0.118 \pm 0.021 \pm 0.026$~\cite{CDFmumuMix};
and in $e$-$\mu$ events
 $\overline{\chi} = 0.130 \pm 0.010 \pm 0.009$~\cite{CDFemuMix}.
All agree with $\overline{\chi} = 0.118 \pm 0.006$
from the PDG~\cite{NewPDG}.

\subsection{
  \Large
Time-Dependent $B^0_d$-Mixing Measurements}
\label{subsec:TimDep}
\begin{figure}[t]
\begin{center}
\epsfxsize=15pc 
\epsfxsize=14pc 
\epsfig{file=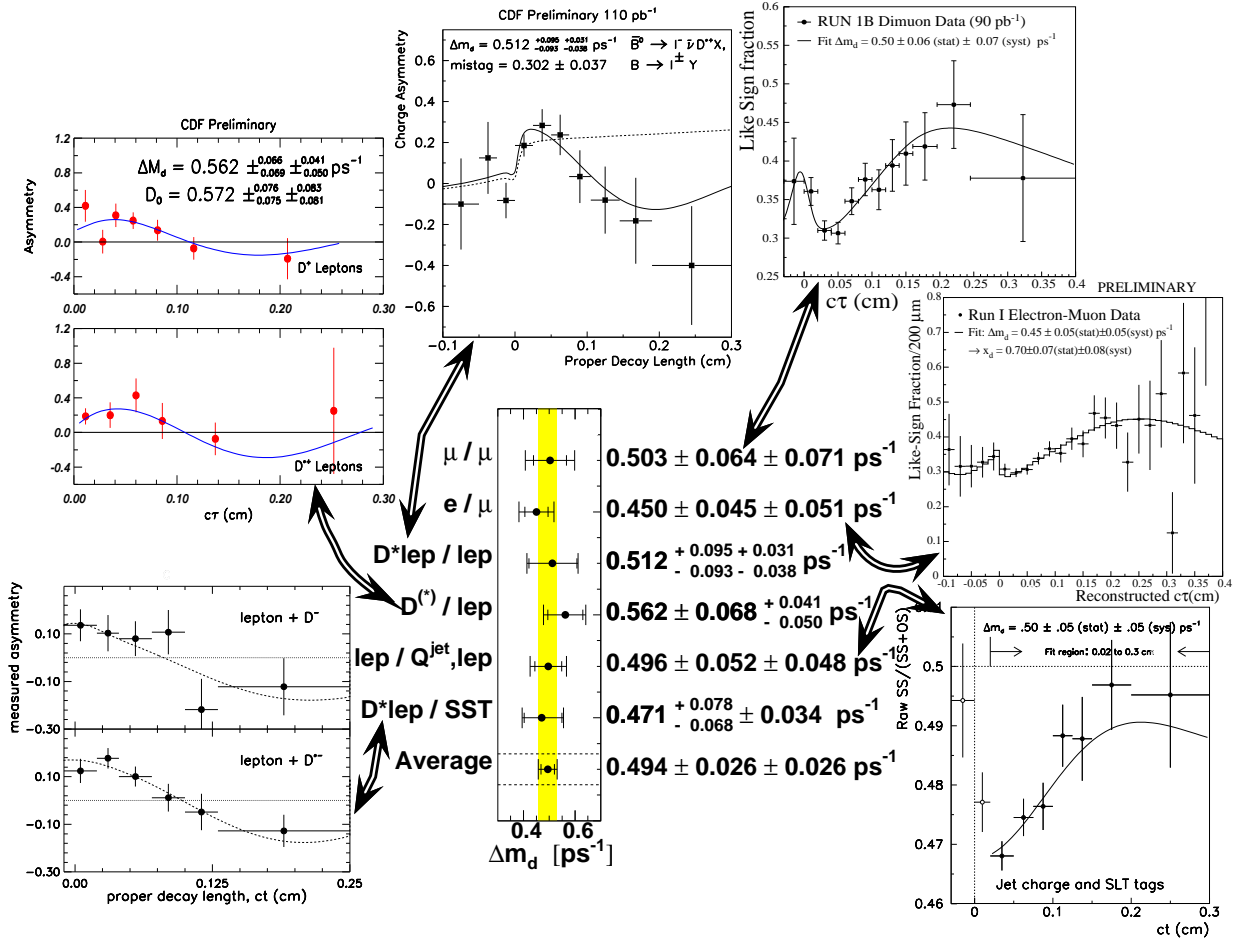,
                width=12.5cm,clip=,angle=-90}
\caption{
Summary of the six CDF $B^0_d$ oscillation measurements
of $\Delta m_d$. In each case
an oscillation is observed in a charge-correlation asymmetry as
a function of proper decay length. The average CDF $\Delta m_d$ 
accounts for correlations between measurements.
\label{fig:CDFdmdSum} 
}
\end{center}
\end{figure}

With the advent of precision vertex detectors
direct observation of the $B^0_d$ oscillation has overshadowed 
the $\overline{\chi}$ analyses.
The D{\O} detector will have such tracking capability starting in
Run II~\cite{D0up},
and such studies have been restricted to CDF in Run I.
Six analyses have been reported ($\sim\!\! 100$ pb$^{-1}$),
and are summarized~in~Fig.~\ref{fig:CDFdmdSum}.

Two analyses are extensions 
of the time-integrated measurements described above: 
dilepton samples ($\sim\!6000\;\mu\mu$, $\sim\!\!10000\; e\mu$)
are used where the leptons define both the $B$ signal and
its flavor. In this case a secondary vertex associated
to a lepton is sought to establish the $B$-decay vertex.
Average $\beta \gamma$ corrections transform
the observed momentum and decay length into the proper decay time.
The inclusive nature of the selection allows other processes to 
contribute ($c\bar{c}$, fakes, etc.\ldots).
Their contributions to the sample
are constrained by kinematic
variables, such as the relative $p_T$ of the lepton to other parts
of the decaying $B$. The samples are more than 80\%  pure $b\bar{b}$. 
The oscillation is revealed in the time variation of the like-sign 
dilepton fraction,
and fits extract $\Delta m_d$ (see ``$\mu/\mu$''~\cite{CDFmumuOsc}
and ``$e/\mu$'' in Fig.~\ref{fig:CDFdmdSum}).

In another dilepton analysis~\cite{tomoko} the $B^0_d$
is more cleanly identified by reconstructing
a $D^*$ near a lepton. From the lepton on the other side one 
infers the initial flavor of the $B \!\rightarrow\! \ell^+ D^{*-} X$ meson.
The  oscillation from a signal of $\sim\!500$ events is shown 
in Fig.~\ref{fig:CDFdmdSum} under ``$D^*$lep/lep,''
along with its $\Delta m_d$.
The cleanliness of the sample results in a small systematic 
error ($^{+0.31}_{-0.38}$), but at the price of worse
statistical precision.

Inclusive lepton triggers ($e$ and $\mu$, $p_T > 8$ GeV/$c$)
are used to obtain a tagged $B$ sample,
where the $B$ flavor on the opposite side is identified 
by reconstructing a $D^{(*)-}$. The resulting oscillation 
for $\sim\!800$ events is
labeled as ``$D^{(*)}$/lep'' in Fig.~\ref{fig:CDFdmdSum}~\cite{stephen}.

We next consider the ``lep/$Q^{jet}$lep'' analysis~\cite{CDFjetqOsc},
which again uses a lepton+vertex to identify a $b$-hadron 
and a second lepton as a tag. It also uses the subtle technique 
of ``jet-charge'' tagging. The sample arises 
from inclusive lepton triggers, 
where we search
for another lepton as a tag
similar to the earlier analyses.
We do not dwell on this aspect.
More interesting is the use of ``jet-charge'' tagging,
where an average charge of a jet opposite the $\ell$+vertex
is used to infer the initial flavor of the $b$-hadron 
producing the trigger lepton.
The charge of the away-side jet is defined as
\begin{equation}
Q_{jet} \equiv \frac{\Sigma_i q_i (\vec{p}_i \cdot \hat{a})}
                    {\Sigma_i      \vec{p}_i \cdot \hat{a} },
\label{eq:JetQ}
\end{equation}
where $q_i$ and $\vec{p}_i$ are the charge and momentum of the $i$-th
track in the jet, and $\hat{a}$ is the unit vector pointing
along the jet axis. A negative (positive) $Q_{jet}$ implies 
the jet 
contained a $b$ ($\bar{b}$). One finds that
the tag purity is higher for larger $|Q_{jet}|$.
The sample composition is again determined via kinematic variables
like the $p_T^{rel}$ of the trigger lepton and the invariant mass
of the secondary vertex. The fit results of the tagging asymmetry
for $\sim\!\frac{1}{4}$ million events are given in Fig.~\ref{fig:CDFdmdSum}
as ``lep/$Q^{jet}$lep.''

The sixth
analysis uses ``Same-Side Tagging'' (SST), 
where a track near the reconstructed $B$ 
tags its flavor~\cite{Gronau}, rather than using the other $b$-hadron.
The idea is simple. 
A $\bar{b}$ quark hadronizing into a $B^0_d$
picks up a $d$ in the fragmentation, 
leaving a $\bar{d}$.
To make a charged pion, the $\bar{d}$
picks up a $u$ making a $\pi^+$. 
Conversely, a $\overline{B}{^0_d}$ will be associated with a $\pi^-$.
Correlated pions also arise from $B^{**+} \rightarrow B^{(*)0} \pi^+$
decays.
Both sources have the same correlation, 
and are not distinguished here.\footnote{
  \normalsize
A recent CDF
$B^{**}$ analysis of a $\ell D^{(*)}$ sample found
the fraction of $B$ mesons arising from
$B^{**}$ states to be $0.28 \pm 0.06 \pm 0.03$~\protect\cite{Dejan}.
$B^{**}$ mesons are indeed a significant source of correlated pions.
}

CDF adopted a ``$p_T^{rel}$'' algorithm, where the tag 
is the candidate track with the smallest momentum component
transverse to the $B$+Track momentum.
A charged particle is a valid SST candidate 
if it is reconstructed in the Si-$\mu$vertex detector (SVX), 
has $p_T > 400$ MeV/$c$,
is within $\Delta R = \sqrt{(\Delta\eta)^2+(\Delta\phi)^2} \le 0.7$
of the $B$,
and its impact parameter is within $3\sigma$ of the primary
vertex.

SST is applied to almost 10,000
$B \rightarrow \ell^+ D^{(*)}X$ events reconstructed via
four $B^0_d$ decay signatures and one for $B^+$~\cite{SSTPRL,SSTPRD}.
The sample composition is unraveled in the fit
(including  $B^0_d \leftrightarrow B^+$  cross-talk).
The $\chi^2$ fit results 
are shown in Fig.~\ref{fig:CDFdmdSum} as ``$D^*$lep/SST,''
with the upper plot showing the $B^0_d \rightarrow \ell^+ {D}{^-}X$
reconstruction and the lower one is for  $\ell^+ \overline{D}{^{*-}}X$.
Along with 
$\Delta m_d$, one also obtains
the dilution,
${\cal D}_0 = 0.181^{+0.036}_{-0.032}$,
 of this SST method.
This analysis provides the dilution calibration
for the SST analyses of Sec.~\ref{sec:cp}.

The six $\Delta m_d$ results are combined, accounting for correlations,
into a CDF average of $0.481 \pm 0.026\pm 0.026$ ps$^{-1}$. 
This is comparable to other experiments, 
and is in good agreement with the PDG value
of $0.464 \pm 0.018$ ps$^{-1}$~\cite{NewPDG}.

\subsection{
  \Large
Time-Dependent $B^0_s$-Mixing Measurements}
$B^0_s$ mixing follows the same formalism
as for $B^0_d$'s, except the relevant CKM element is $V_{ts}$
rather than $V_{td}$.
Measurements of $\overline{\chi}$ (Sec~\ref{sec:timeint}), 
along with $\chi_d$ measured at the $\Upsilon(4S)$
and estimates for the species fractions $f_d$ and $f_s$, 
provided the original direct indication that $\chi_s$
is close to its asymptotic limit ($1/2$) and thereby
insensitive to $\Delta m_s$.
Further progress on $B^0_s$ mixing necessitates
time-dependent methods.

\begin{figure}[t]
{\centering
  \epsfxsize=13pc 
  \epsfxsize=18pc 
\epsfbox{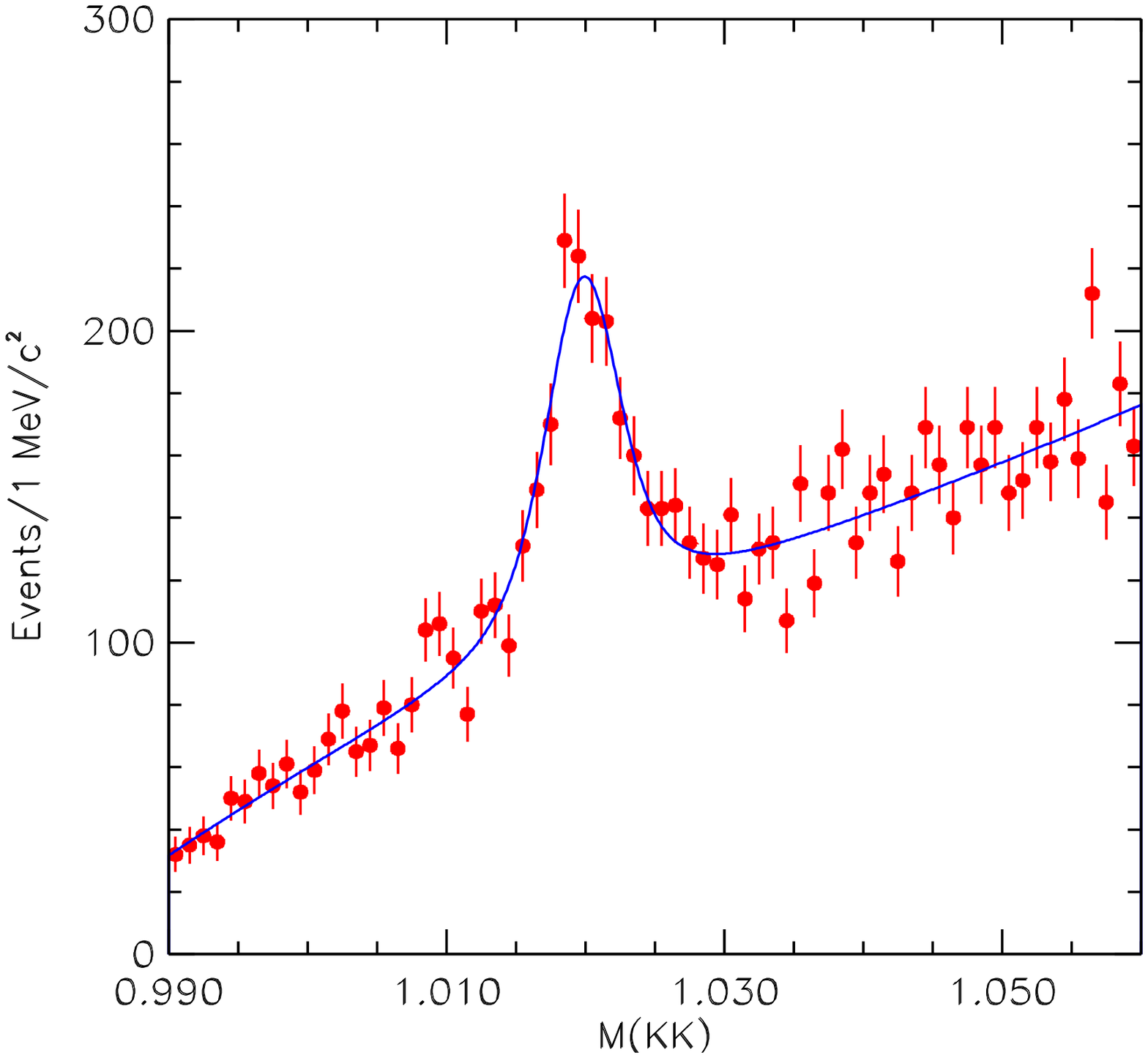}
  \epsfxsize=18pc 
\epsfbox{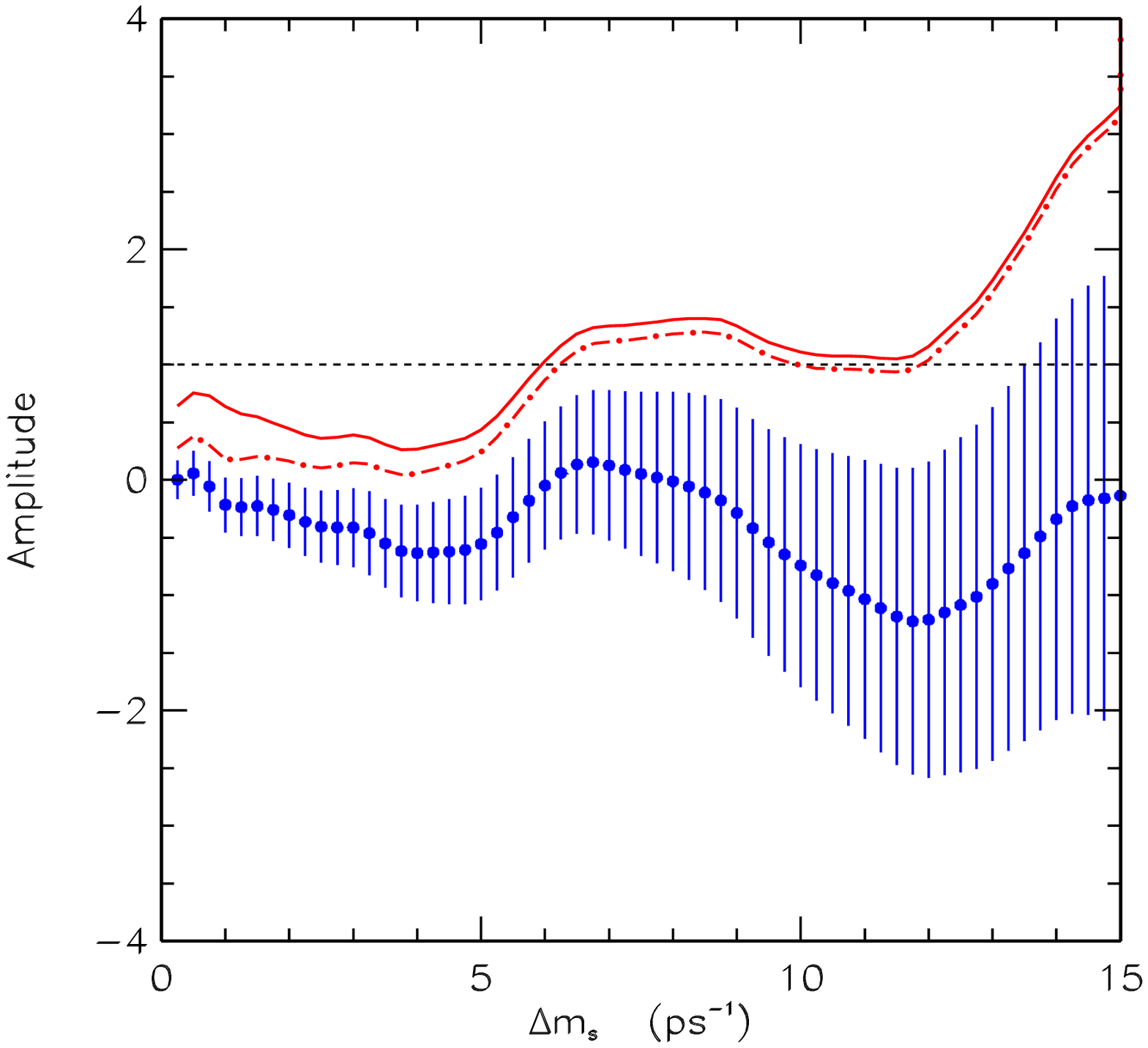}
\caption{
  \normalsize
         Left: The $K^+K^-$ mass distribution showing the $\phi$ peak.
         Right: The likelihood amplitude scan through $\Delta m_s$ (points),
         and the 95\% CL limits on the amplitude for statistical error
         (dashed curve) and statistical and systematic errors combined
         (solid curve).
\label{fig:Bs} 
}}
\end{figure}

CDF has searched for $B^0_s$ oscillations~\cite{CDFBs} using 
dileptons (110 pb$^{-1}$ of $\mu\mu$ or $e \mu$) 
for $\ell^\pm\ell^+ \phi h^-$, where $h^-$ is a charged track
associated to a $\phi \rightarrow K^+K^-$ decay vertex. 
This signature selects $B^0_s \rightarrow \ell^+ \nu D^-_s$,
$D^-_s \rightarrow \phi \pi^- X$; exclusive reconstruction
is not required to increase statistics.
The resulting $K^+K^-$ mass distribution is shown in Fig.~\ref{fig:Bs}.
The $B^0_s$ decay point is obtained by projecting the $\phi h^-$ 
decay back to the $\ell^+$. 
Monte Carlo corrections are applied to the $\beta\gamma$ factor 
for the proper time estimation.
A sample of $1068 \pm 70$ $B^0_s$ candidates
(purity of $61.0 ^{+4.4}_{-7.0}$\%) is obtained.

The $B^0_s$ flavor is inferred from the other trigger lepton,
similar to the $\Delta m_d$ analyses.
Events are again classified as ``unmixed'' ($\ell^\pm\ell^\mp_{tag}$)
or ``mixed'' ($\ell^\pm\ell^\pm_{tag}$).
Limits will be set rather than an observation of the 
oscillation, so one must know {\it a priori} the mistag rate $R_{m}$.
This was found to be $R_{m} = \frac{1}{2}(1-{\cal D}_0) = 0.24 \pm 0.08$ 
from a likelihood fit of the mixed/unmixed fractions 
in the $\ell^+ \phi$ data.
The $B^0_s$ oscillation is too rapid to influence 
the determination of $R_{m}$;
rather it is governed 
by the sample contributions of $B^0_d$, $B^+$, $c\bar{c}$, 
sequential decays and fake background.

The data are fit with an unbinned likelihood to describe 
the mixed versus unmixed components. 
The fit includes the various sources
of events ($B^0_s$, $B^0_d$, $B^+$,\ldots) and $\Delta m_s$ is
a free parameter.
No oscillation is observed, and limits are set. 
The amplitude method~\cite{HGMoser} is adopted, whereby the functional form
of $\cos(\Delta m_s t)$ is replaced by $A(\Delta m_s)\cos(\Delta m_s t)$,
{\it i.e.} the amplitude is a free $\Delta m_s$-dependent
parameter. For the true value $\Delta m_s^{true}$, 
$A(\Delta m_s^{true}) = 1$, and otherwise $A(\Delta m_s) = 0$.
The result of the scan of $A(\Delta m_s)$ from the likelihood 
is shown in Fig.~\ref{fig:Bs}. The data fluctuate
about zero, with no evidence for an oscillation.
Values of $\Delta m_s$ are excluded at the 95\% CL if 
$A(\Delta m_s) + 1.645\sigma_A(\Delta m_s) \le 1$,
and thus $\Delta m_s > 5.8$ ps$^{-1}$ at 95\% CL
accounting for both statistical and systematic errors
(Fig.~\ref{fig:Bs}, solid line).

This result is competitive with other single tagging measurements.
However, the world limit, $\Delta m_s > 12.4$ ps$^{-1}$ (95\% CL),
is dominated by the multi-tag results from
ALEPH and DELPHI~\cite{Parodi}.

It is conceivable that $B^0_s$ oscillations are too rapid 
to be directly observed. 
If so, the width difference $\Delta \Gamma_s$ between the
mass eigenstates is expected to be large.
CDF has searched for two lifetime components in a $\ell D_s$
sample, and finds $\Delta \Gamma_s/\Gamma_s < 0.83$ at 95\% 
CL~\cite{CDFBsLife}.
Given $\Delta \Gamma_s/\Delta m_s$ and the mean $B^0_s$ lifetime
$\overline{\tau}_s$,
this can be expressed as the upper bound
$\Delta m_s < 96 \, {\rm ps}^{-1} \times 
[5.6 \times 10^{-3}/(\Delta \Gamma_s /\Delta m_s)] 
[1.55 {\rm ps}/\overline{\tau}_s]$ at 95\% CL.,
with $\Delta \Gamma_s /\Delta m_s = 5.6 \times 10^{-3}$
a recent estimate~\cite{Beneke}.
The limit is weak, but with the increased statistics 
of Run II either $\Delta m_s$ or $\Delta \Gamma_s$ should be
directly determined.

\section{
  \Large
 $CP$ Violation in $J/\psi K^0_S$}
\label{sec:cp}
The origin of $CP$ violation has been an outstanding question
since its unexpected discovery 
in $K^0_L \rightarrow \pi^+\pi^-$ 35 years ago~\cite{KL_CP}.
In 1972, before the discovery of charm,
 Kobayashi and Maskawa~\cite{CKM} proposed that this was 
the result of quark mixing with 3 (or more) generations.
Unfortunately the $K^0$ has been the only place $CP$ violation
has been observed. Despite precision $K^0$-studies,
a complete picture of $CP$ violation
is still lacking; and it is often argued
that the CKM model can {\it not} be the full story~\cite{Cosmo}.

$CP$-violation searches have encompassed $B$ mesons,
but the  effects in
inclusive studies~\cite{CDFmumuMix,B_CP}  
are too small ($\sim\!\!10^{-3}$) to as yet detect.
In the early '80's it was realized~\cite{CarterSanda}
that the  mixing  {\it interference} of $B^0_d$
decays into the same $CP$ state could manifest large violations.
Unfortunately these 
 decays were, until recently, 
too rare to study.

The  ``golden'' mode for observing large $CP$ violation in $B$'s
is $B^0_d/\overline{B}{^0_d} \rightarrow J/\psi K^0_S$,\footnote{
  \normalsize
Another
mode of possible interest is $B^0_d \rightarrow J/\psi K^{*0}$.
While this is not a $CP$ eigenstate, it can be
decomposed into even and odd eigenstates by an angular analysis. 
This has been done by CDF for
$B^0_d \rightarrow J/\psi K^{*0}$ (and
$B^0_s \rightarrow J/\psi \phi$) to extract the decay matrix 
elements~\cite{Pappas}, but the Run~I statistics are insufficient
to be of interest 
for measuring $CP$-violation parameters.
}
and, critically, it is related to the CKM matrix
with little theoretical uncertainty. 
A $B^0_d$ may decay directly to $J/\psi K^0_S$,
or the $B^0_d$ may oscillate into $\overline{B}{^0_d}$
and then decay to $J/\psi K^0_S$. These two paths have
a phase difference, quantified by the angle $\beta$ 
of the unitarity triangle (Fig.~\ref{fig:Unitarity1}). 
This gives rise to a decay asymmetry 
\begin{equation}
{\cal A}_{CP}(t) \equiv \frac{\overline{B}{^0_d}(t)-B^0_d(t) }
                        {\overline{B}{^0_d}(t)+B^0_d(t) } = 
                       \sin(2\beta) \sin (\Delta m_d t),
\label{eq:cp_asym}
\end{equation}
where $B^0_d(t)$ [$\overline{B}{^0_d}(t)$] is the number of decays
to $J/\psi K^0_S$ at proper time $t$ given that the 
meson was a $B^0_d$ [$\overline{B}^0_d$] at $t=0$.
OPAL investigated $CP$ violation with 
24 $J/\psi K^0_S$ candidates (60\% purity),
and obtained $\sin(2\beta) = 3.2^{+1.8}_{-2.0} \pm0.5$~\cite{OPALcp}.
CDF has taken advantage of the large $B$ cross section at the
Tevatron and obtained a sample of several hundred decays to $J/\psi K^0_S$
to measure $\sin(2\beta)$~\cite{CDFcp}.

\subsection{
  \Large
Same Side Tagging Analysis of $J/\psi K^0_S$\protect}
\label{sec:SSTcp}
\begin{figure}[t]
\begin{center}
\epsfxsize=17.5pc 
\epsfxsize=22pc 
\epsfbox{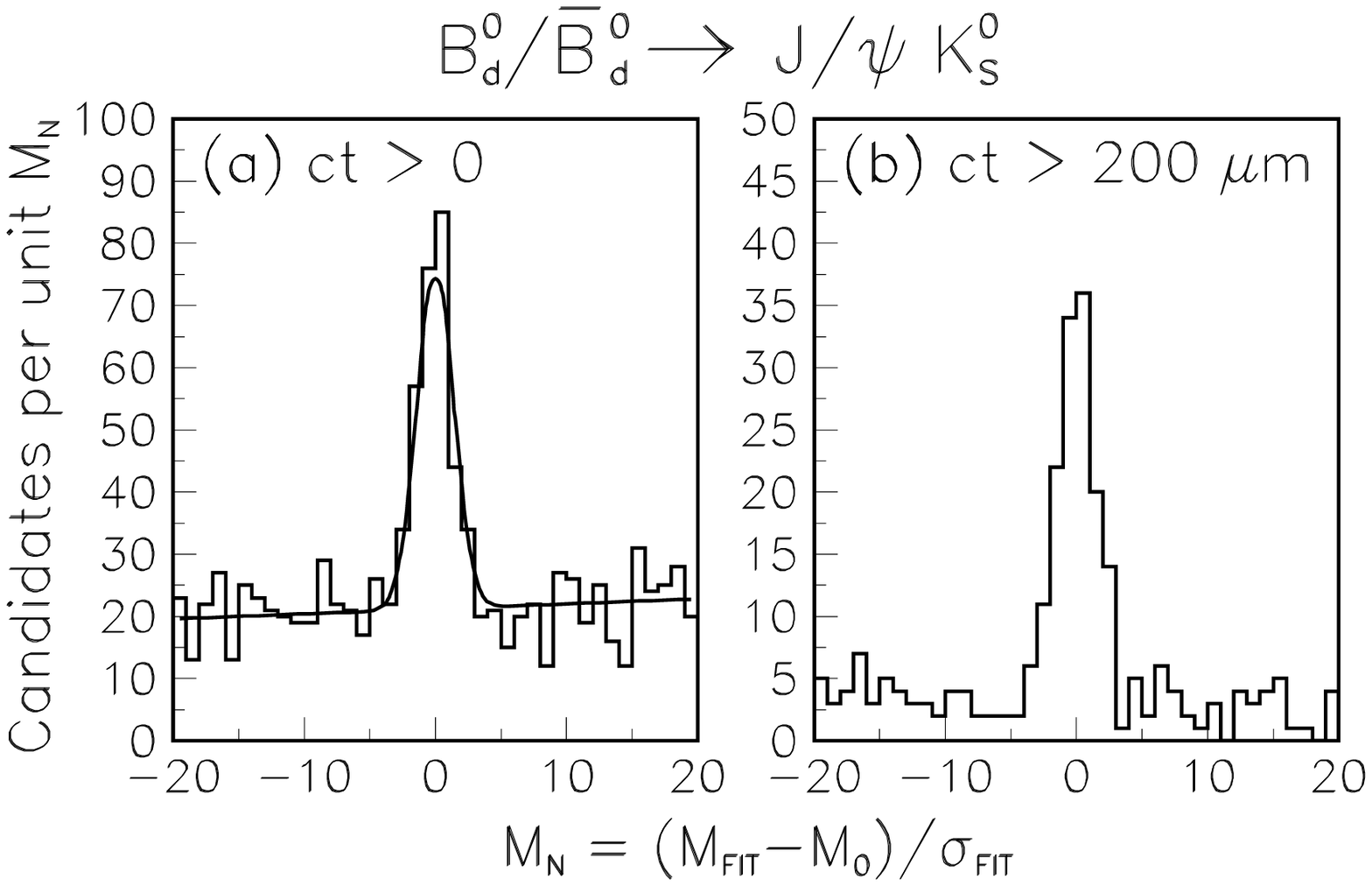}
\epsfxsize=12.0pc 
\epsfxsize=16pc 
\epsfbox{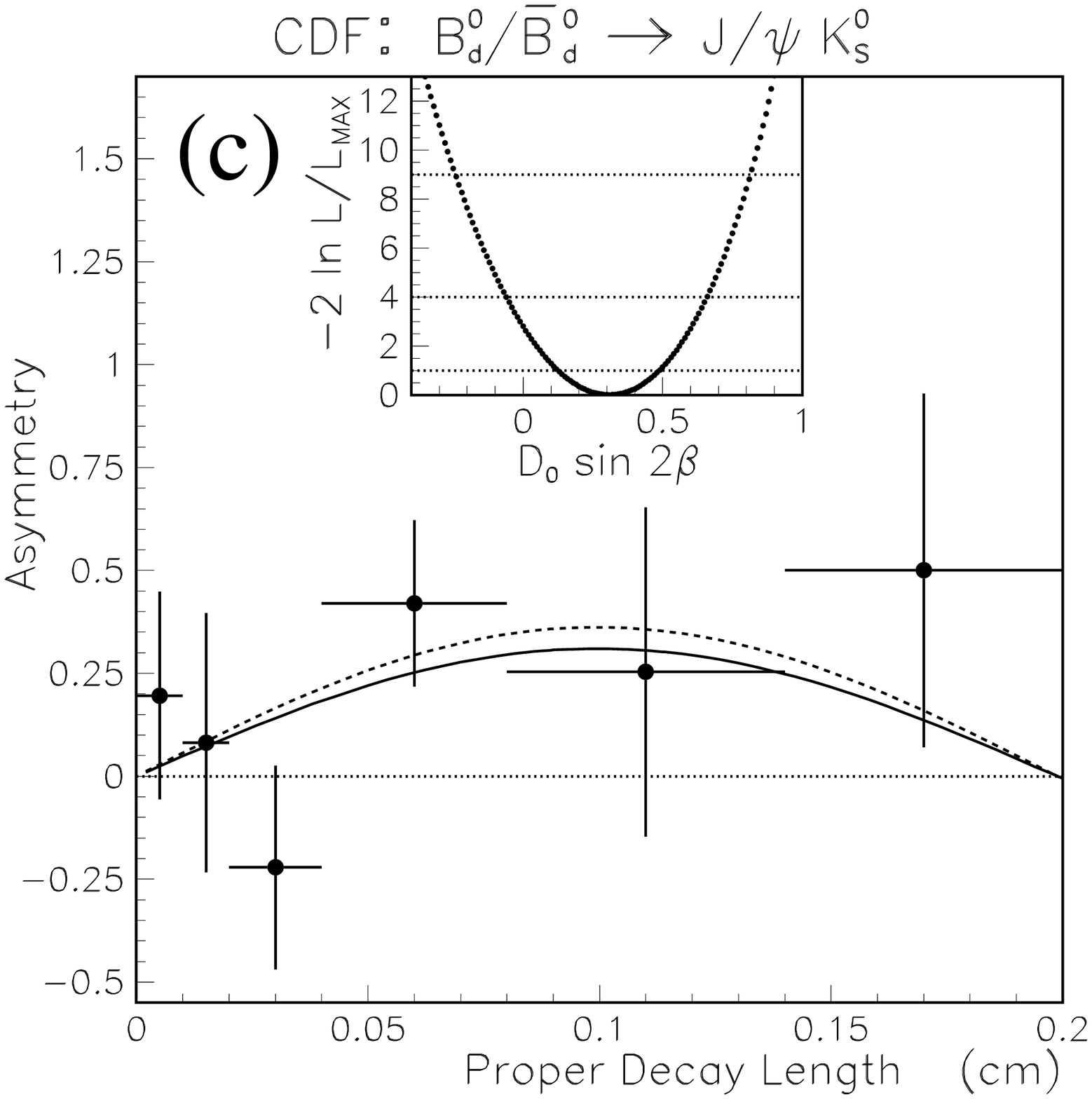}
\caption{
  \normalsize
The normalized mass distribution of $J/\psi K^0_S$ 
events with $ct > 0$ (a) and $200 \, \mu$m (b). 
The curve is the Gaussian signal plus linear background 
from the likelihood fit (see text).
The sideband-subtracted flavor asymmetry (c)
as a function of the $J/\psi K^{0}_s$ proper decay length  
[Eq.~(\ref{eq:meas_asym})]:  points are data, 
dashed curve is a simple $\chi^2$ fit, 
and the solid curve is the likelihood fit.
The inset shows a scan of the log-likelihood function 
as ${\cal D}_0\sin(2\beta)$ is varied about the best fit value.
\label{fig:JpsiKs} 
}
\end{center}
\end{figure}

$J/\psi K^0_S$ candidates are selected 
from the $J/\psi \rightarrow \mu^+\mu^-$ sample  (110 pb$^{-1}$).
A time-dependent analysis demands that
both muons are in the Si-$\mu$vertex detector (SVX), 
cutting away half of the $J/\psi$'s.
The $K^0_S \rightarrow \pi^+\pi^-$ reconstruction tries all
tracks, assumed to be pions.
The $p_T(K^0_S)$ must be above 0.7 GeV/$c$,
its decay vertex displaced 
from the $J/\psi$'s by more than $5\sigma$,
and $p_T(B^0_d) > 4.5$ GeV/$c$.
 We construct $M_N \equiv (M_{FIT} - M_0)/\sigma_{FIT}$, where
 $M_{FIT}$ is the fitted $J/\psi K^0_S$ 
 mass, $\sigma_{FIT}$ its error ($\sim\!\!9\;{\rm MeV}/c^2$), and
 $M_0$ is the central  $B^0_d$ mass.
The distribution for candidates with $ct >0$
is shown in Fig.~\ref{fig:JpsiKs}a.
A likelihood fit yields (for all $ct$) $198 \pm 17$ $B^0_d/\overline{B}{^0_d}$'s.

The initial flavor is tagged by
the identical SST of the $\ell D^{(*)}$ $\Delta m_d$
measurement of Sec.~\ref{subsec:TimDep}.
SST is independent of the $B^0_d$ decay mode,
and therefore the dilution measurement can be transferred from
one mode to another.
However a small kinematic correction, determined from Monte
Carlo, is made to translate
the $\ell D^{(*)}$ dilution to the  $J/\psi K^0_S$ sample
due to the different $p_T(B)$ ranges~\cite{SSTPRD}.
The appropriate ${\cal D}_0$ is $0.166 \pm 0.018 \pm 0.013$~\cite{CDFcp},
where the first error is due to the dilution measurements,
and the second is due to the translation to the $J/\psi K^0_S$ sample.

The SST method is applied to the $J/\psi K^0_S$ sample,
with a resultant tagging efficiency of $\sim\!65\%$.
Analogous to Eq.~(\ref{eq:cp_asym}), we compute the asymmetry 
\begin{equation}
{\cal A}(ct) \equiv \frac{N^-(ct)-N^+(ct) }
                         {N^-(ct)+N^+(ct) },
\label{eq:meas_asym}
\end{equation}
where $N^\pm(ct)$ are the numbers of positive and negative tags 
(implying $B^0_d$ and $\overline{B}{^0_d}$ respectively)
in a given $ct$-bin. Signal and sideband
regions are defined as $|M_N|<3$ and $3<|M_N|<20$, and
the sideband-subtracted asymmetry of Eq.~(\ref{eq:meas_asym})
is plotted in Fig.~\ref{fig:JpsiKs}c.
The dashed curve is a $\chi^2$ fit of ${\cal D}_0 \sin(2\beta)\sin(\Delta m_d t$) 
to the data, with $\Delta m_d$ fixed to 
$0.474 \;{\rm ps}^{-1}$~\cite{PDG1}.
The amplitude, $0.36\pm 0.19$, 
measures $\sin(2\beta)$ attenuated by the dilution ${\cal D}_0$.
Due to the $\sin(\Delta m_d t)$ shape
the fit amplitude is driven by the asymmetries
at larger $ct$'s, where backgrounds are small (see Fig.~\ref{fig:JpsiKs}b).

The fit is refined
using an unbinned likelihood fit.
This makes optimal use of the low 
statistics by fitting the data in $M_N$ and $ct$,
including sideband and $ct<0$ events which help constrain the background.
The fit also incorporates resolutions and
corrections for (small) systematic detector biases.
The solid curves in Fig.~\ref{fig:JpsiKs}a,c are the result of the likelihood
fit, which gives ${\cal D}_0\sin(2\beta)=0.31\pm 0.18$. As expected,
both fits give similar values since the result is 
dominated by the sample size.
The systematic uncertainty on ${\cal D}_0\sin(2\beta)$ is 0.03,
dominated by the uncertainty on $\Delta m_d$ ($\pm 0.031$ ps$^{-1}$), 
but includes the effects from
detector biases and the $B^0_d$ lifetime.

To extract $\sin(2\beta)$ from the measured asymmetry the dilution
must be divided out. 
However, as long as ${\cal D}_0 \!\not = \! 0$, the exclusion 
of $\sin(2\beta)\!=\! 0$ is {\it independent} of further knowledge
of ${\cal D}_0$. Given ${\cal D}_0 > 0$,
the unified frequentist approach of Feldman and Cousins~\cite{Feldman}
yields a dilution-independent
exclusion of $\sin(2\beta) \le 0$ at 90\%  CL.

From above, ${\cal D}_0 = 0.166 \pm 0.018 \pm 0.013$,
which results in $\sin(2\beta)= 1.8 \pm 1.1 \pm 0.3$.
The central value is unphysical since the amplitude of the raw
asymmetry is larger than ${\cal D}_0$.
This result corresponds to excluding  $\sin(2\beta)<-0.20$
at a 95\% CL.

\subsection{
  \large
Multi-Tagging Analysis of $J/\psi K^0_S$}
Shortly after this conference CDF released an expanded
$J/\psi K^0_S$ analysis using three taggers,
as well as non-SVX data.
These preliminary results are briefly summarized.

The above SST result is statistically very limited. It
can be improved by increasing the {\it effective} statistics 
by using lepton and jet-charge tagging to increase the total
$\epsilon {\cal D}^2_0$.
The raw statistics can also be increased by utilizing the $J/\psi$ candidates
not reconstructed in the SVX. Precision lifetime information
is lost, reducing the power of these events, but significant
information remains.
The $J/\psi K^0_S$ selection is otherwise similar to the SST-only
analysis,
and results in a total of $395 \pm 31$  $J/\psi K^0_S$ candidates
($202 \pm 18$ in the SVX, $193 \pm 26$ are non-SVX).

SST, lepton, and jet-charge tagging are applied to this sample.
Lepton tagging
follows the $\Delta m_d$ analyses. 
The jet-charge algorithm is similar to the  $\Delta m_d$ analysis
but uses a ``mass''
jet algorithm rather than a ``cone'' based one. This improves the
efficiency for identifying low-$p_T$ ``jets'' for the tag.
A lepton tends to dominate the jet 
charge if a lepton tag
is in the jet.
Since lepton tagging has low efficiency
but high dilution, the correlation between lepton and 
jet-charge tags is avoided by dropping the jet-charge tag
if there is a lepton tag.
The dilutions of these two methods are measured  
in $B^+ \rightarrow J/\psi K^+$ decays (Table~1),
and are directly applicable to the $J/\psi K^0_S$ sample.
The precision is not high, but this obviates the complex problem
of translating dilutions from kinematically different 
samples. 

The tagged $J/\psi K^0_S$ events are fit in an unbinned likelihood fit
($\Delta m_d$ constrained 
to $0.464 \pm 0.018$ ps$^{-1}$~\cite{NewPDG}).
The $\sin(2\beta)$ results for the individually tagged subsamples
are listed in Table~1,
 \begin{table}[t]
 \caption{
 \normalsize 
Multi-tag analysis of $J/\psi K^0_S$.
There are two dilutions for SST, one for SVX events and
another for non-SVX events.}
 \label{tab:multi}
 \begin{center}
\begin{tabular}{|c|c|c|c|}
\hline
Tagger        &Eff. (\%)&Dil. (\%)& $\sin(2\beta)$ \\      \hline
\raisebox{0pt}[10pt][3pt]{$\!$SST$_{svx}\!$}
                        & $\!35.5\pm 3.7\!$ & $\!16.6 \pm 2.2\!$ &\\
\raisebox{0pt}[3pt][5pt]{$\!$SST$_{non}\!$}
                        & $\!38.1 \pm 3.9\!$ & $\!17.4 \pm 3.6\!$  
                        & \raisebox{5pt}[7pt][0pt]{$\!\;2.03^{+0.84}_{-0.77}$}\\ 
\raisebox{0pt}[7pt][6pt]{$\!$Lepton$\!$}   
                        & $\!5.6  \pm 1.8\!$ &$\!62.5 \pm 14.6\!$ 
                        &  $\!\;0.52^{+0.61}_{-0.75} \!$ \\
\raisebox{0pt}[7pt][7pt]{$\!$Jet-$Q\!$}    
                        & $\!40.2 \pm 3.9\!$ &$\!23.5 \pm 6.9\!$  
                        & $\!\!-0.31^{+0.81}_{-0.85}\! $ \\ \hline
\raisebox{0pt}[11pt][6pt]{$\!$Global$\!$}
                        & \multicolumn{2}{|c|}{$\epsilon {\cal D}^2 = 6.3\pm1.7$}
                        & $\!\;0.79 ^{+0.41}_{-0.44}\!$ \\ \hline 
\end{tabular}
 \end{center}
 \end{table}
 \begin{figure}[t]
  \centering
  \protect\epsfxsize=22pc 
\hspace*{-0.2cm}\protect\epsfbox{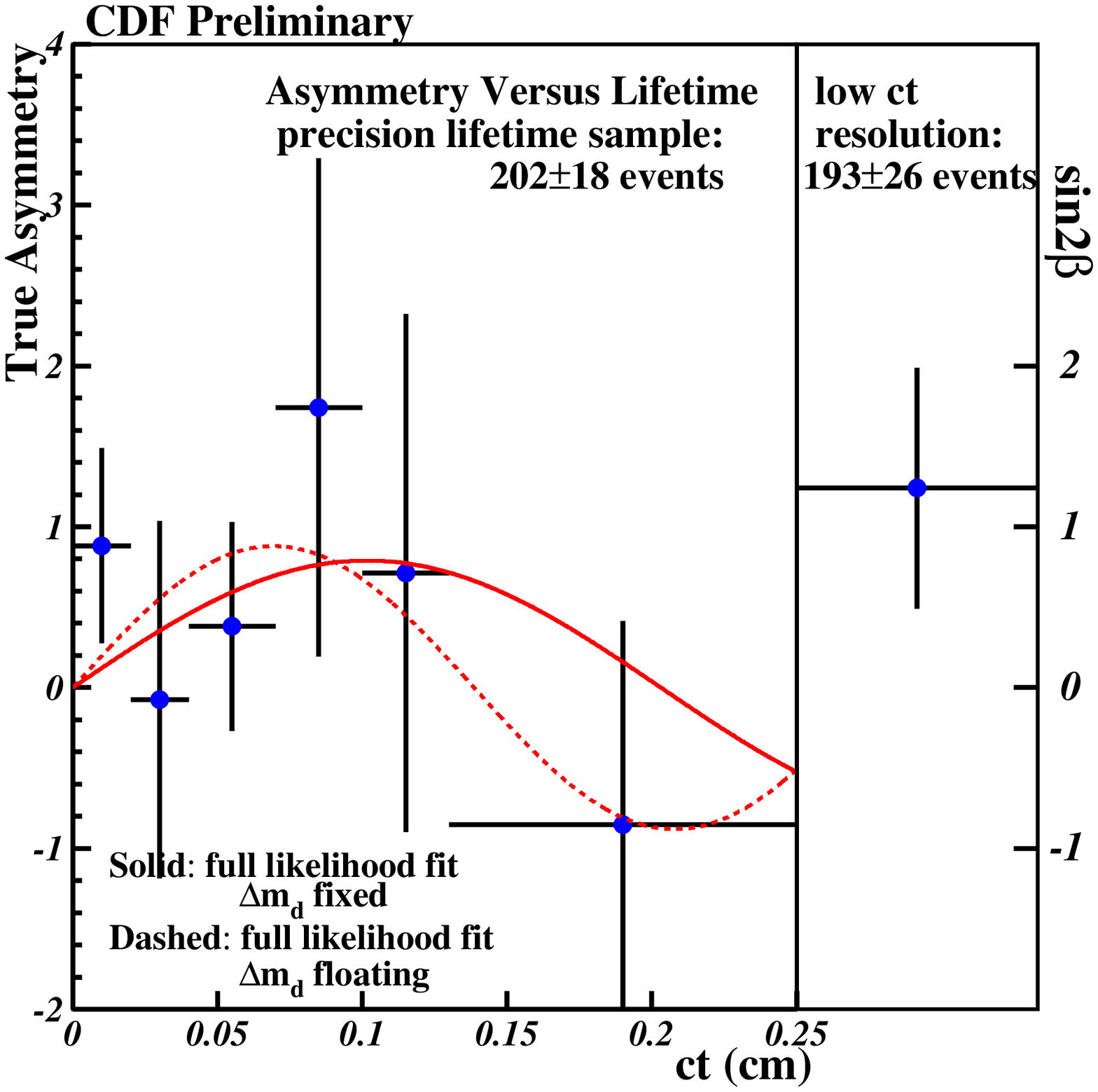}
\protect\epsfxsize=12pc 
  \caption{
  \normalsize
Multi-tag $\sin(2\beta)$ result. 
Left: time-dependent asym\-metry of SVX data;
Right: time-integrated asym\-metry of non-SVX data.
  }
 \protect\end{figure}
including systematic errors
due to the dilutions, $\Delta m_d$, $\tau_{B^0}$, and $m_{B^0}$.
The SST result is slightly larger than before with the inclusion of the
non-SVX events, and the error has decreased by $\sim\!\!20$\%.
The other two taggers fall in the physical range, one positive
and the other negative.

Rather than average these three results, the likelihood fitter
is generalized to fit all three simultaneously
while accounting for tag correlations. 
The global multi-tag result is 
$\sin(2\beta) = 0.79^{+0.41}_{-0.44}$ (Fig.~5),
including the systematic uncertainties.
This result corresponds to a unified frequentist confidence
interval of $0.0 < \sin(2\beta) < 1$ at 93\% CL.
Although the exclusion of zero has only slightly increased 
from 90\% for the (SVX) SST-only analysis, 
the uncertainty on $\sin(2\beta)$ is cut in half.

\section{
  \Large
Summary and Prospects}
\begin{figure}[t]
\begin{center}
\epsfxsize=15.5pc 
  \epsfxsize=20pc
\epsfbox{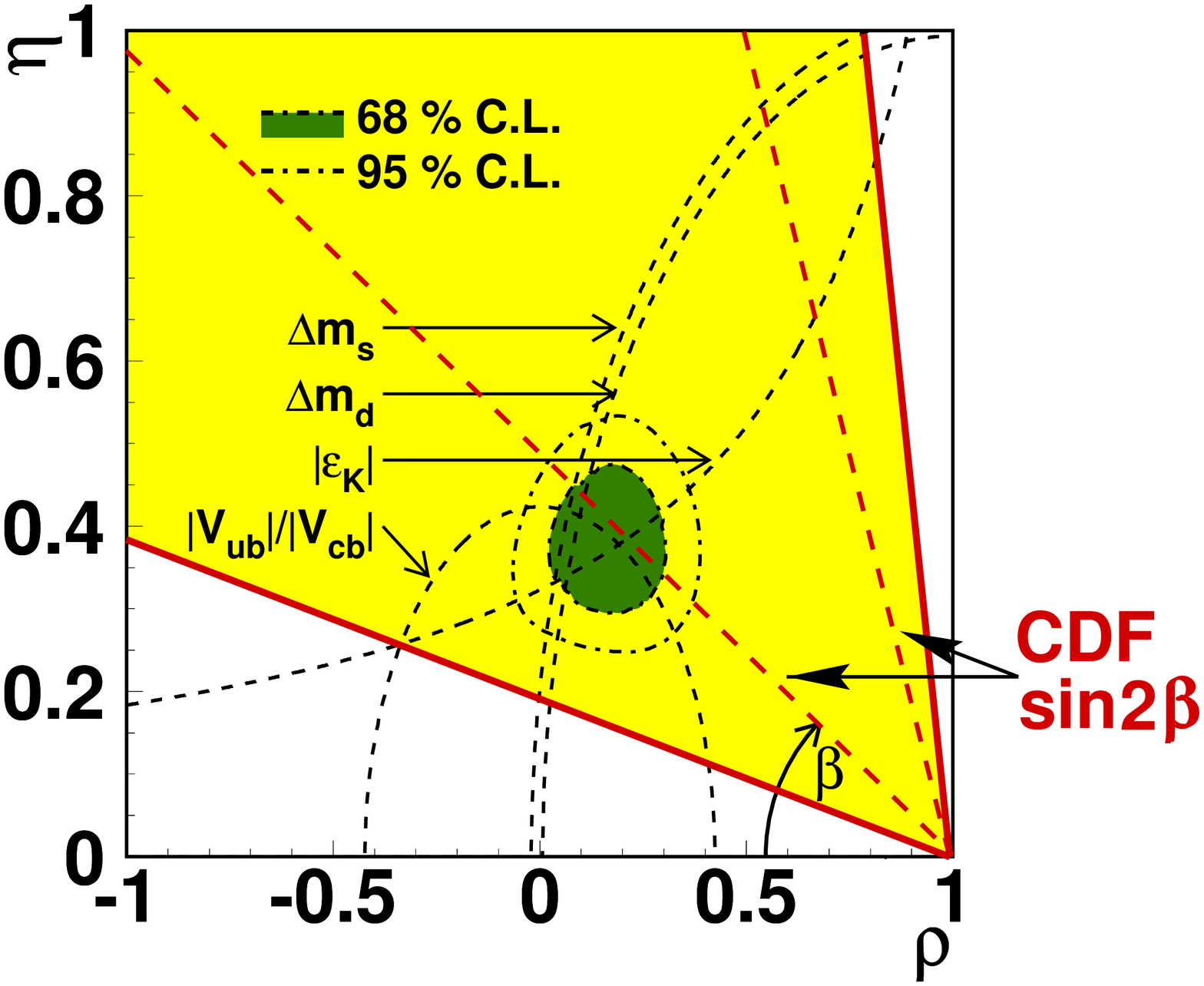}
\epsfxsize=13pc 
  \epsfxsize=16pc
\epsfbox{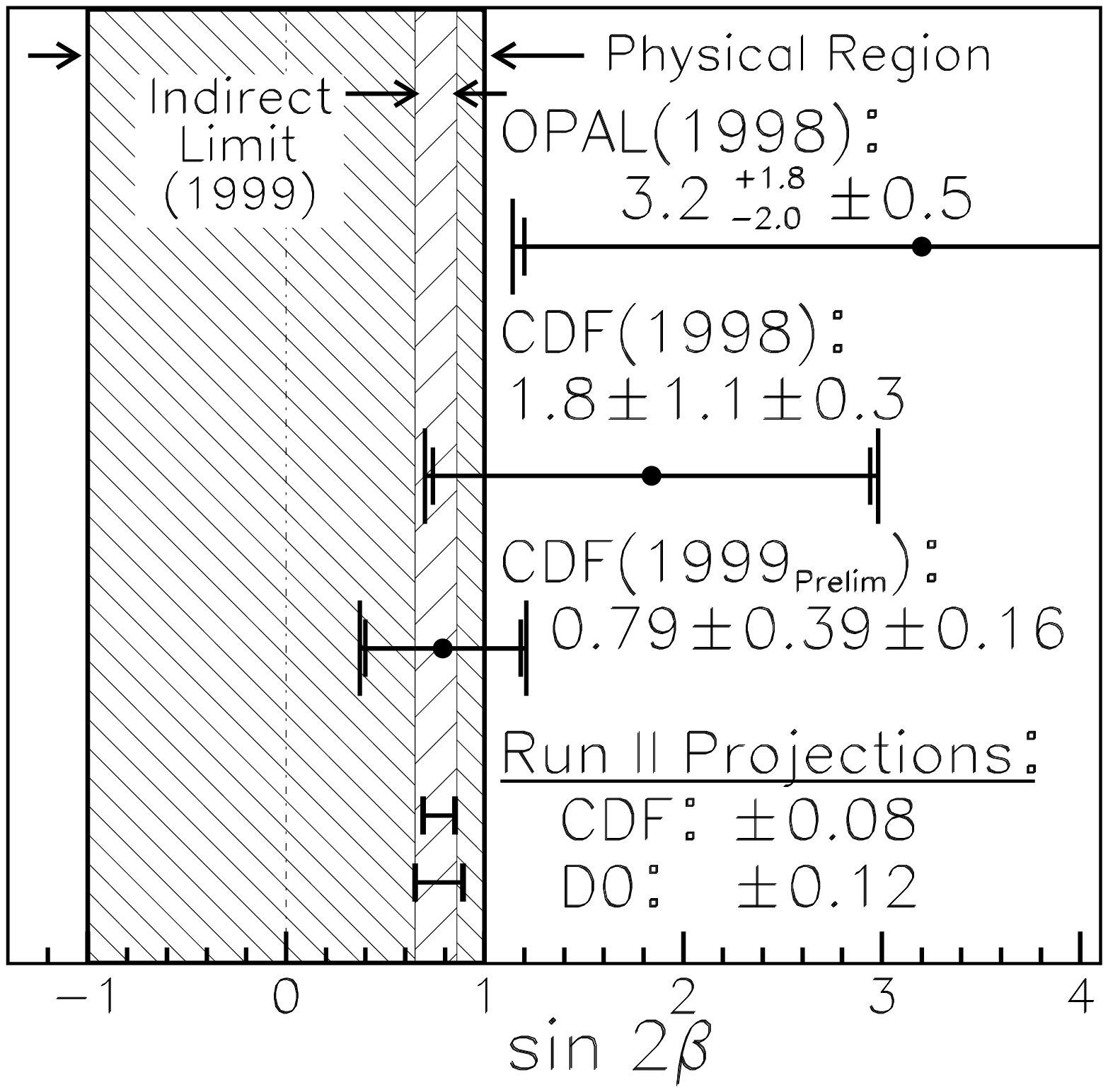}
\vspace*{-0.4cm}
\caption{
  \normalsize
Left: CKM constraints from 
$K^0_L$'s, charmless $B$ decays,
$\Delta m_d$, and $\Delta m_s$ on ($\rho,\eta$).
The ``$1\sigma$'' allowed regions of Fig.~\protect\ref{fig:Unitarity1}
are collapsed to their central values, except for $\Delta m_s$
which is a 95\% CL limit boundary excluding the left region.
The combined constraints on ($\rho,\eta$) are indicated 
by the 68 and 95\% CL contours, and translate
to  $\sin(2\beta) = 0.75 \pm 0.09$~\protect\cite{Mele}.
Superimposed is the CDF $\sin(2\beta)$ measurement
with two of the four solutions for $\beta$ shown (long dashed lines).
The $1\sigma$ allowed range (light shaded region) 
is shown for the smaller $\beta$ solution.
Right: Summary of $\sin(2\beta)$ measurements~\protect\cite{OPALcp,CDFcp},
Run II projections, and an indirect ``$1\sigma$'' range reported 
at this Conference~\protect\cite{Ali}.
\label{fig:summary} 
}
\end{center}
\end{figure}

Measurements of $\Delta m_{d}$ and $\Delta m_{s}$ provide important
constraints on the CKM matrix.
World averages constrain the triangle 
of Fig.~\ref{fig:Unitarity1} quite well,
as shown in Fig.~\ref{fig:summary}.
An indirect determination
of $\sin(2\beta) = 0.77 ^{+0.09}_{-0.12}$
was reported at this conference~\cite{Ali}.
This is much more precise than the direct CDF measurement;
nevertheless, the agreement is an auspicious 
omen for the CKM model's account of $CP$ violation.

Additional sources of 
$CP$ violation are, however,  thought necessary
to account for the baryon asymmetry in the universe~\cite{Cosmo}.
Searching for physics beyond the CKM model 
demands stringent tests, and
is the focus of dedicated $B$ factories.
Both CDF and D{\O} will also be fully engaged in this effort 
by exploiting the rich $B$ harvest from Run II. 
Commencing in 2000,
a two-year run
will deliver $20\times$ the luminosity ($\sim\!\!2$~fb$^{-1}$),
to be exploited by greatly enhanced detectors.
With the ``baseline'' detector and
trigger upgrades~\cite{D0up,CDFup} CDF 
projects 10,000 $J/\psi K^0_S$
dimuon triggers for a $\sin(2\beta)$ error of about $\pm 0.08$;
and D{\O}, with a new precision tracking system,
expects an error of 0.12-0.15 (Fig.~\ref{fig:summary}).
Dielectron triggers may further increase the samples by  $\sim\!50$\%.
These errors are in the range projected for $B$ factories.
The Run~II Tevatron will be competitive in many other areas
of $B^0_d$ physics as well.

$B^0_s$ oscillations have, so far, eluded all comers.
The Tevatron should have a virtual monopoly on the $B^0_s$
after the closure of the $Z^0$ machines and before the start of the LHC.
Run II baseline expectations are for CDF 
to reach $x_s$'s up to 30-40, and D{\O} up to  20-25.
Although D{\O}'s reach is less, it is sufficient that failure 
to observe oscillations 
should critically challenge the CKM model.

In addition to the baseline upgrades, both experiments are aggressively
pursuing further improvements. For example, 
CDF is working towards an additional
Si-layer to improve vertex resolution and a Time-of-Flight system,
which may push  $x_s$  out to $\sim\!60$;
and D{\O} is looking at a displaced-track trigger to
greatly enhance $B$ triggering.

We close with the CKM model unscathed, 
but look forward to an exciting future where,
perhaps, some of the mystery surrounding $CP$ violation may
be unveiled.

\section*{
  \Large
Acknowledgments}
I would like to thank my fellow collaborators, and
 colleagues across the ring, for the pleasure 
 of representing them and their work.
Assistance with this presentation from
T.~Miao,
M.~Paulini,
C.~Paus,
F.~Stichelbaut,
and
J.~Tseng
is appreciated.

\end{document}